\documentclass[a4paper,11pt]{article}
\usepackage{pos}
\usepackage{subcaption}
\usepackage{microtype}
\usepackage{caption}
\usepackage{titlesec}
\titlespacing{\section}{0pt}{7pt}{4pt}

\usepackage[utf8]{inputenc}

\title{The Terzina Payload on NUSES: Silicon Photomultipliers Performance and Radiation Damage Mitigation in Low Earth Orbit}
\ShortTitle{SiPM Performance and Radiation Damage Mitigation for Terzina}

\author*[a]{Shideh Davarpanah}
\onbehalf{for the \textsc{NUSES} Collaboration}  

\affiliation[a]{Département de Physique Nucléaire et Corpusculaire, Université de Genève, Faculté de Sciences, \\1211 Genève, Switzerland}

\emailAdd{shideh.davarpanah@unige.ch}

\abstract{
Silicon photomultipliers (SiPMs) are increasingly favored for detecting near-UV, visible, and infrared light in space due to their high sensitivity to single photons and compact design. While SiPMs offer several advantages over traditional photomultiplier tubes, their susceptibility to radiation and noise remains an issue. Our study intended to determine the optimal SiPM design for the Terzina Cherenkov satellite in sun-synchronous low Earth orbit (LEO) at an altitude of 550~km. To this end, we characterised several NUV-HD-MT SiPM variants, developed by Fondazione Bruno Kessler, and studied their responses to irradiation by exposing them to a 50~MeV proton beam and to electrons from the $\beta$-emitting source strontium-90. Experimental results were cross-validated with SPENVIS and Geant4 simulations. Based on our findings, we also propose an annealing strategy to mitigate radiation damage and ensure reliable performance in space-based applications.}

\ConferenceLogo{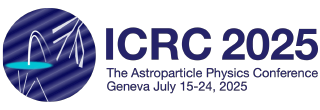}

\FullConference{39th International Cosmic Ray Conference (ICRC2025)\\
 15–24 July 2025\\
Geneva, Switzerland\\}
\begin{document}
\maketitle

\section{Introduction: Silicon Photosensors in Space Applications}

Silicon Photomultipliers (SiPMs) are increasingly favored for space-based applications due to their lightweight design, low power consumption, and high photon detection efficiency (PDE). Their compact and robust structure makes them particularly suitable for integration into space instruments. However, their operation in space is challenged by two main factors: sensitivity to temperature fluctuations and radiation exposure. Both effects can significantly increase the already considerable dark count rate (DCR). While extreme or rapidly varying temperatures may cause short-term performance variations, long-term exposure to radiation results in a gradual degradation of the device.

Several current and upcoming space missions are leveraging SiPM technology to advance our understanding of high-energy astrophysical phenomena. For example, SiPMs produced by Hamamatsu Photonics are foreseen for use in the plastic scintillator detector of the HERD (High Energy Cosmic Radiation Detection) mission~\cite{HERD}. Another mission employing scintillating crystals read out by SiPMs is Crystal Eye~\cite{Barbato_2019}. The work presented here focuses on addressing these limitations to improve the reliability and performance of SiPMs for space applications and their implementation in the NUSES mission.

This proceeding is organized as follows: Section~\ref{sec:terz_desc} provides an overview of the NUSES mission and, in particular, its Terzina Cherenkov telescope, including its design and scientific objectives. Section~\ref{sec:sipm_char} presents the characterization of SiPM performance, covering static, dynamic, and optical response. In Section~\ref{sec:rad_study}, we discuss the results of proton and electron irradiation campaigns, evaluate the thermal dependence of SiPM parameters, model the evolution of the DCR during the mission, and propose an annealing-based mitigation strategy. Finally, Section~\ref{sec:conclusion} summarizes our findings and their implications for future space-based SiPM applications.

\begin{figure}[t!]
    \centering      
    \includegraphics[width=0.5\textwidth]{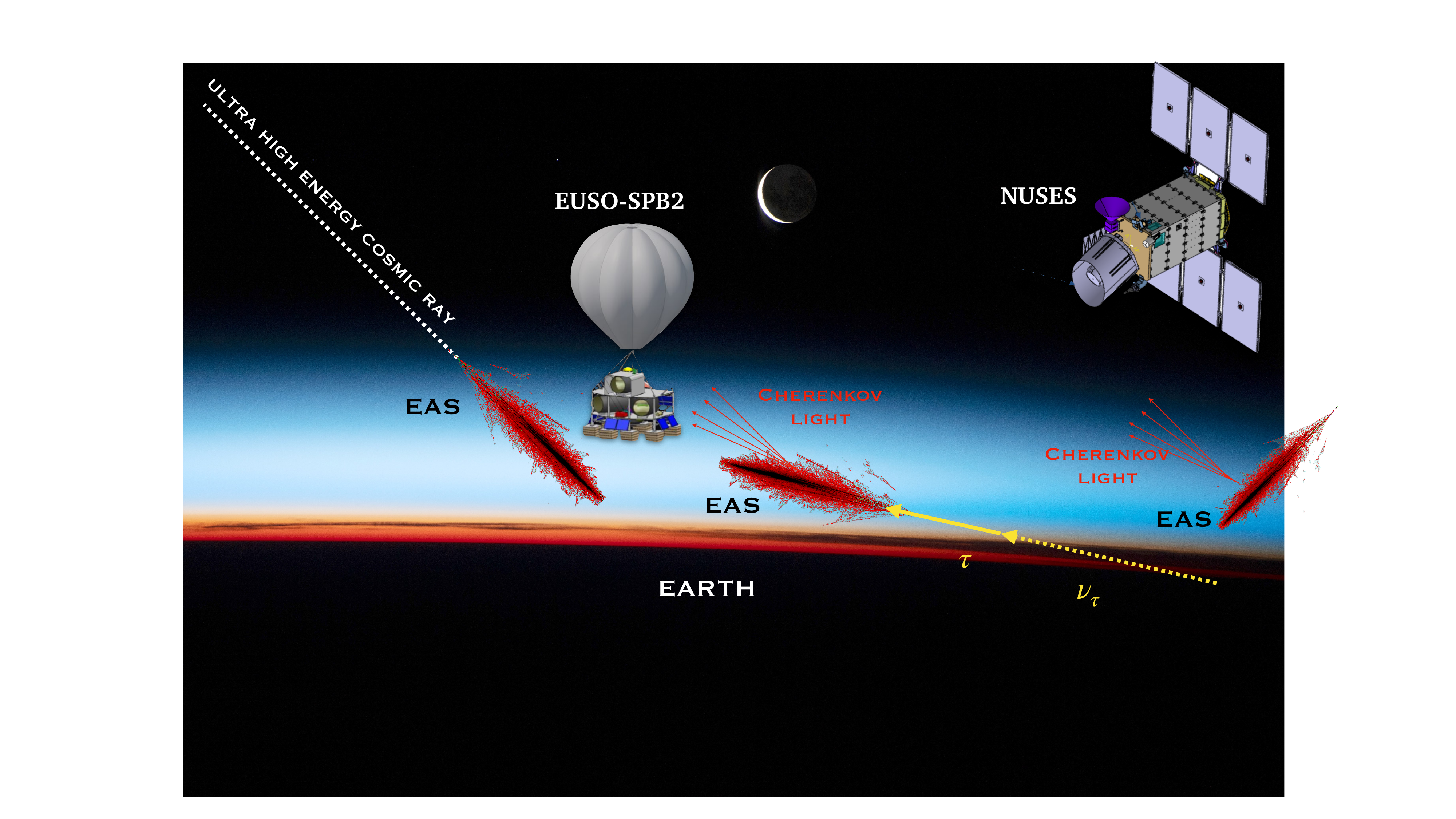}
        \caption{\label{fig:NUSES} Pictorial view of extensive air showers detection by space-based ones like Terzina onboard NUSES.}
\end{figure}

\section{The Terzina telescope onboard NUSES}
\label{sec:terz_desc}

The NUSES~\cite{Trimarelli:2024rdb} mission is a space-based astroparticle physics project that will operate in a quasi-polar low Earth orbit (LEO). It carries two scientific payloads: Terzina\cite{Burmistrov_2023} and Zirè\cite{DeMitri:2023ejv}, both of which utilize SiPMs for photon detection. Zirè is designed to monitor low-energy cosmic rays ($<300$MeV), gamma-ray bursts in the energy region 0.1-50~MeV, and space weather phenomena, using SiPMs coupled with scintillating crystals. In contrast, Terzina focuses on the detection of ultra-high-energy cosmic rays (UHECRs) and Earth-skimming neutrinos through the observation of Cherenkov light generated by extensive air showers (see Fig.~\ref{fig:NUSES}). In addition to its scientific objectives, NUSES serves as a pathfinder satellite, testing innovative technologies for future space-based astroparticle physics detectors. 

The satellite platform, developed by the industrial partner TAS-I (Thales Alenia Space – Italia), will operate at altitudes ranging between 550~km (beginning-of-life) and 535~km (end-of-life). The Terzina payload is a near-UV telescope optimized for the detection of faint Cherenkov light. It employs a Schmidt-Cassegrain optical system, characterized by a dual-mirror design with an effective focal length of 925~mm. The primary mirror has a diameter of 444~mm, while the secondary aspherical mirror, 201.6~mm in diameter, is paired with a 30~mm-thick correcting lens. The overall effective collecting area is approximately 0.11~m$^2$, with about 8\% optical shadowing caused by mechanical structures such as vanes and baffles. The final mechanical and optical systems were designed and are being implemented by Officina Stellare, following specifications provided by the University of Geneva and the Gran Sasso Science Institute.

\begin{figure}[t!]
    \centering
    \includegraphics[height=3.8cm]{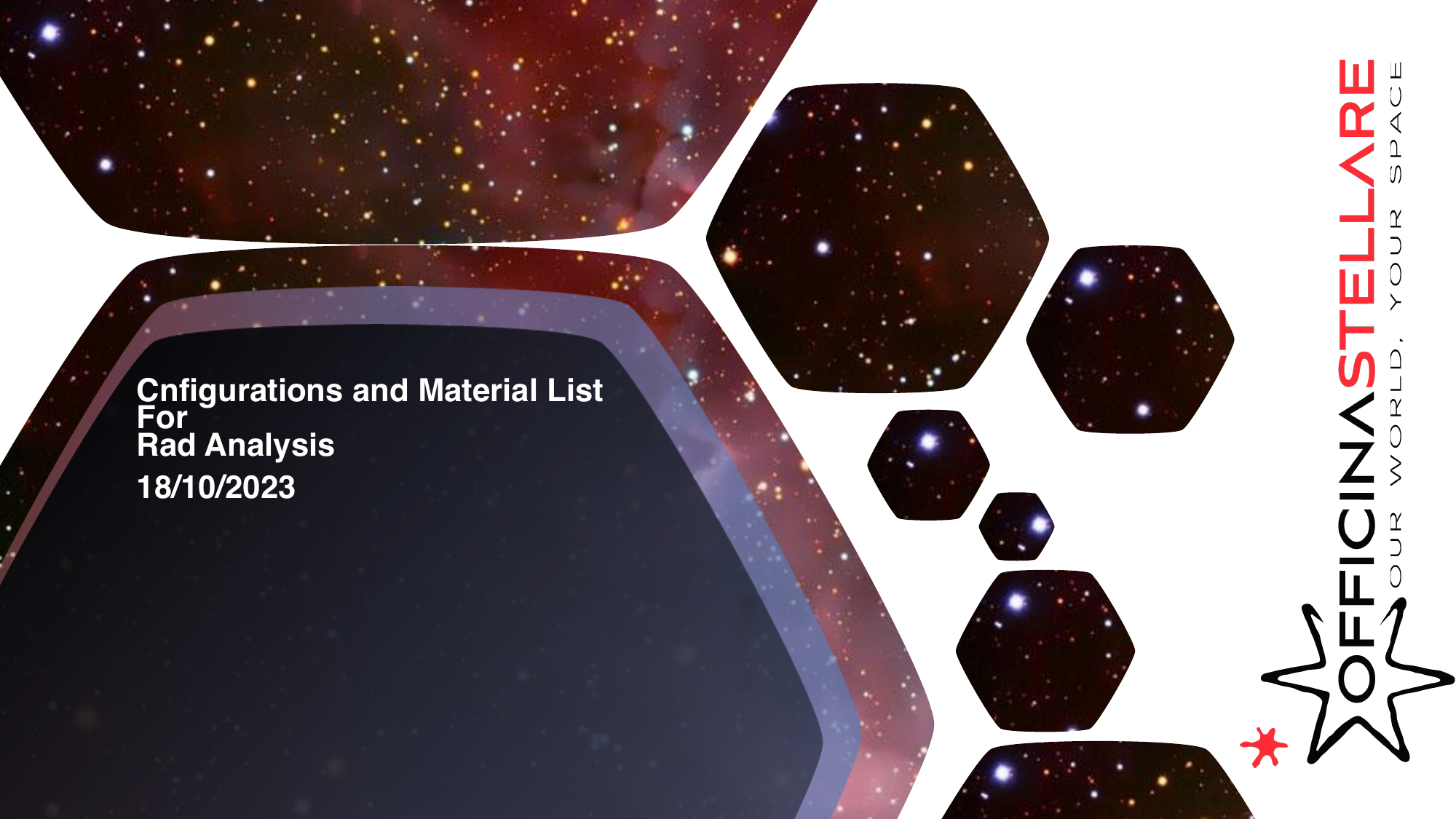}
    \includegraphics[height=3.8cm]{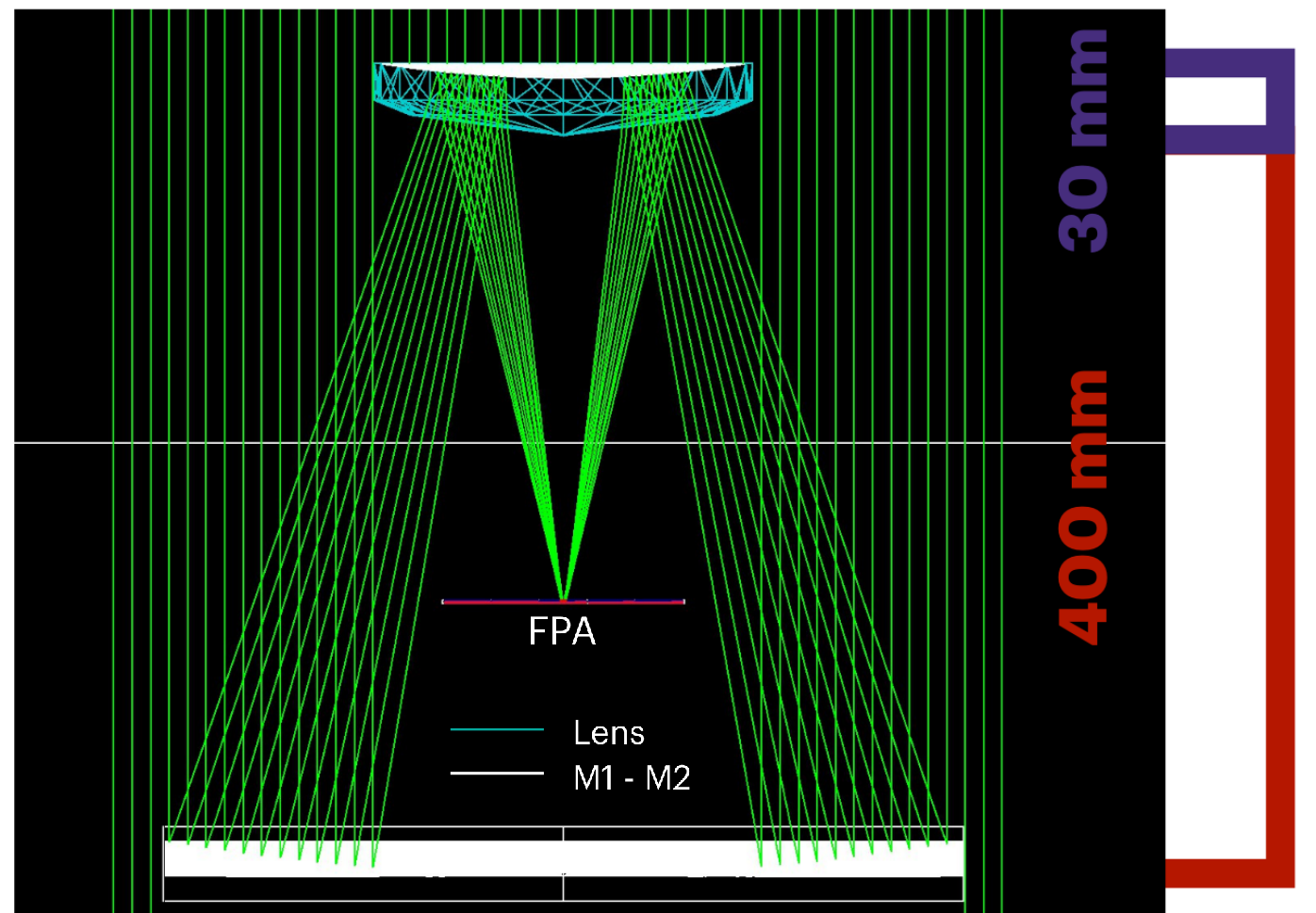}
    \caption{Left panel: Side view of the preliminary geometry of the telescope.
    Right panel: Preliminary optical layout of the Terzina telescope with the Schmidt-Cassegrain optics.} 
    \label{fig:terz_desc}
\end{figure}

The focal plane assembly (FPA) consists of ten SiPM tiles arranged in two rows of five. Each tile contains an $8\times8$ array of $3\times3$~mm$^2$ pixels, for a total of 640 pixels. Each pixel has a sensitive area of approximately $2.900\times2.625$~mm$^2$ and covers a field of view (FoV) of about $0.18^\circ$. This results in a total FoV of $7.2^\circ$ horizontally and $2.56^\circ$ vertically. Due to the image inversion inherent in Cassegrain optics, the upper row of the FPA is pointed toward below the Earth’s limb and is intended for background and atmospheric light monitoring, as well as potential Earth-skimming neutrino events. The lower row targets above the limb for UHECR-induced extensive air showers (EAS). Detailed mechanical and optical schematics of the system are shown in Fig.~\ref{fig:terz_desc}.

\section{Characterization of SiPM Performance}
\label{sec:sipm_char}

The FPA employs NUV-HD-MT~\cite{s19020308} SiPMs from Fondazione Bruno Kessler (FBK), which feature deep trench isolation (DTI) filled with metal to suppress optical cross-talk, maintaining it below 10\%~\cite{gola2022} at up to 10~V overvoltage. To find the best micro-cell sizes that maximized the photon detection efficiency (PDE) while keeping the signal duration short,
%The signal duration is proportional to the size and capacitance of the micro-cells. A large signal increases the noise captured within the acquisition window required to integrate the full pulse. 
we have investigated NUV-HD-MT SiPMs of $3\times3$~mm$^2$ and $1\times1$~mm$^2$ dimensions with square micro-cell sizes of  25, 30, 35, 40, 50~$\mu \text{m}$. A further significant consideration is whether to utilise bare or coated sensors. While bare sensors require meticulous handling, with coated sensors, electrons may produce Cherenkov light within the resin layer. All laboratory measurements were carried out at the University of Geneva and at room temperature ($T=22$~$ ^{\circ}$C), following the setup and analysis methodology described in~\cite{our_papaer, Nagai_2019}.
%in~\cite{Nagai:2019yzb,Nagai_2019}. A detailed, step-by-step description of the experimental setup and measurement procedures can be found in~\cite{our_papaer}.

\begin{figure}[t!]
    \centering
    \includegraphics[height=4.5cm]{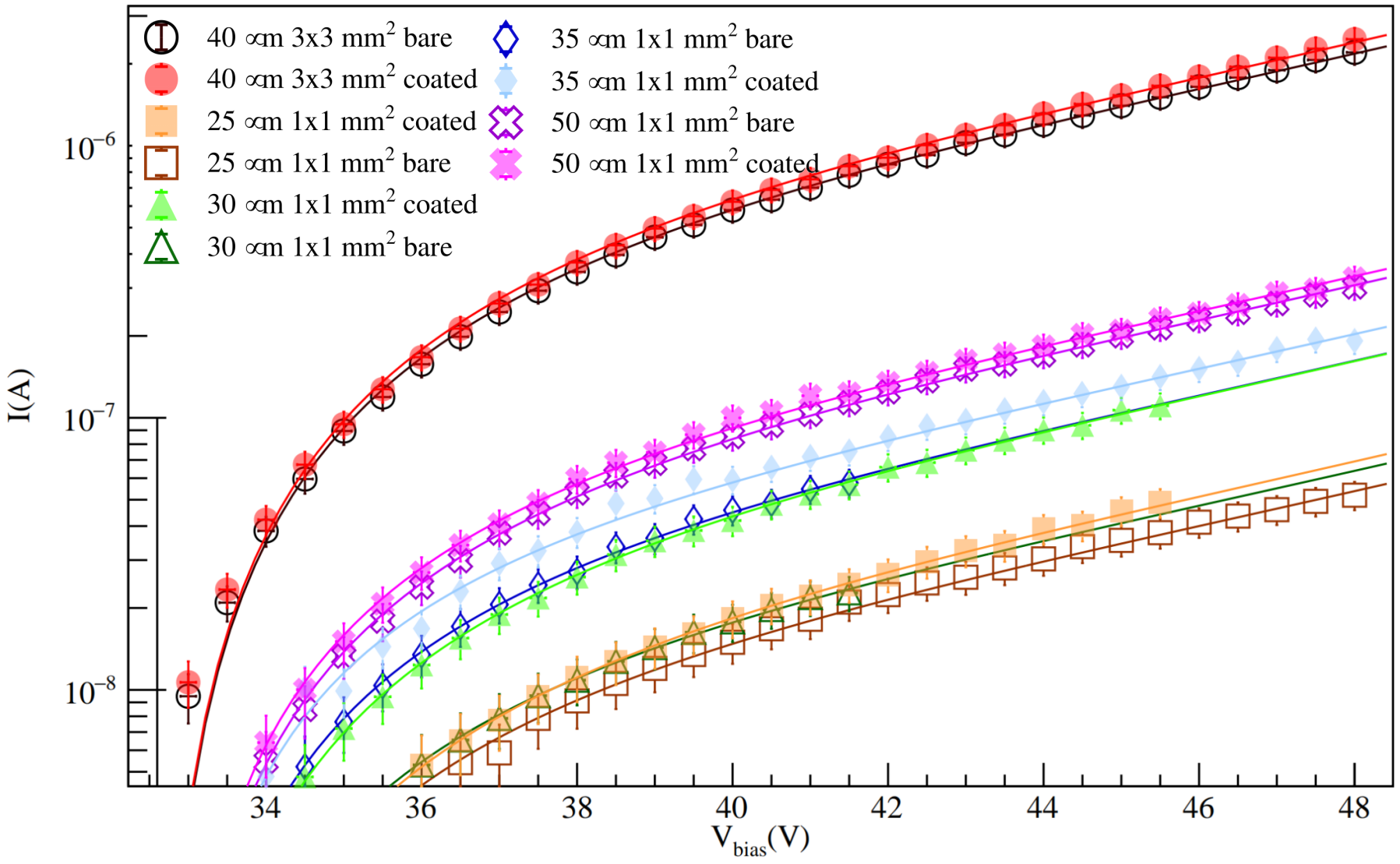}
    \includegraphics[height=4.5cm]{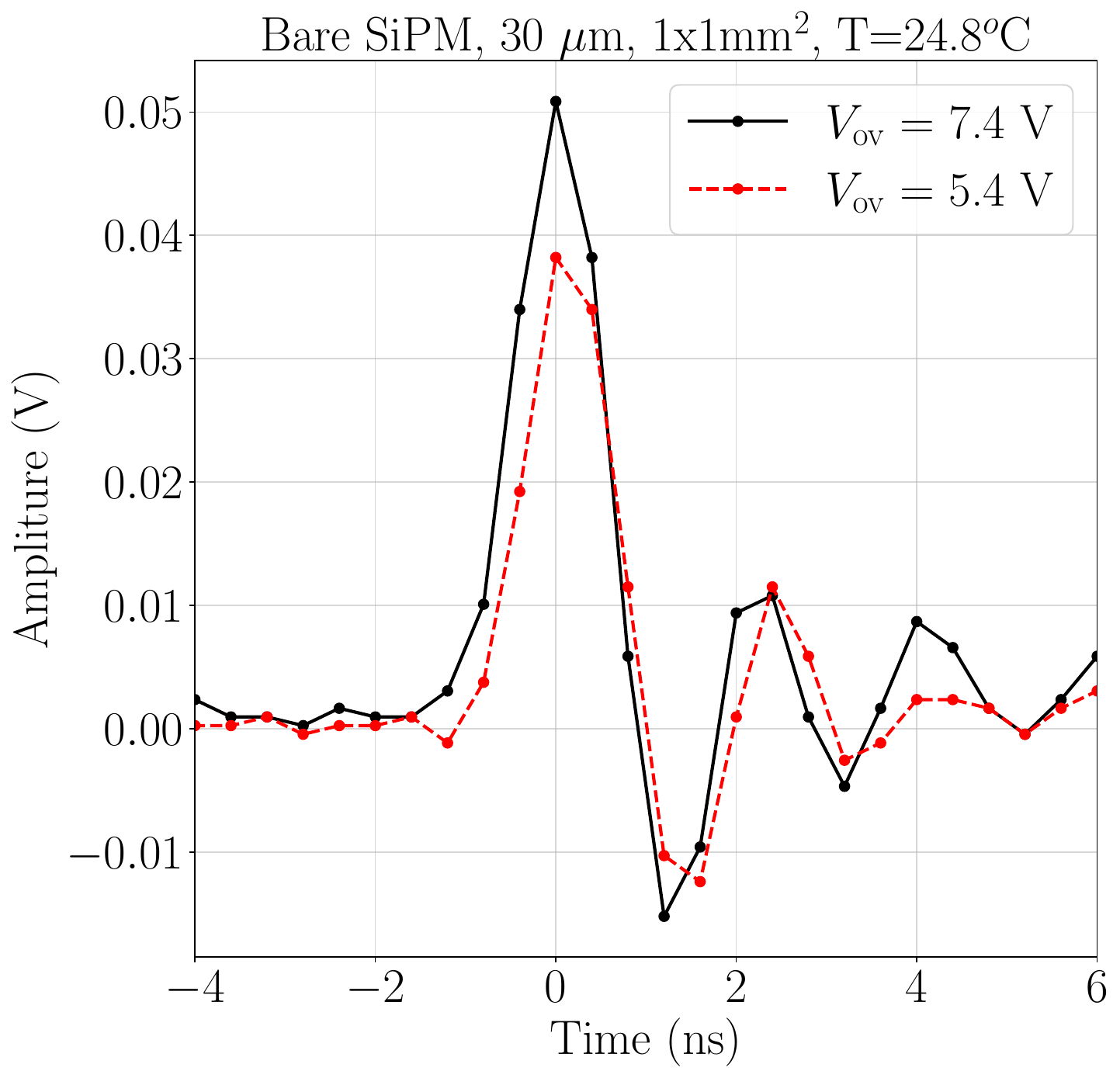}
    \caption{Left panel: The reverse IV characteristics (data symbols) and model (lines) for various SiPM samples at room temperature. Right Panel: Single photoelectron response after the preamplifier for different overvoltages for a bare sensor with 30~$\mu$m micro-cell size.}
    \label{fig:IV_postbreakdown_fit}
\end{figure}

%\subsection*{Static Characterization} % Static characterization
\textbf{Static Characterization:} We performed the static characterisation of the SiPMs by analysing their forward and reverse IV curves to extract key parameters such as quenching resistance ($R_q$) and breakdown voltage ($V_{\mathrm{BD}}$). We found $R_q \sim 1$~M$\Omega$ for both bare and coated sensors of the same size but with varying micro-cell dimensions. Using the second log-derivative method, we found that $V_{\mathrm{BD}}$ is approximately 32.6~V at room temperature across all tested devices. To describe the post-breakdown regime—dominated by Geiger-mode avalanche processes—we employed an analytical ``IV model"~\cite{Dinu:2016hog,Nagai_2019} that incorporates carrier-induced current, micro-cell capacitance, and the Geiger probability. This approach accurately captures the SiPM behaviour above the breakdown voltage. Fig.~\ref{fig:IV_postbreakdown_fit} shows the fit of this model for reverse IV data. 
%The model shows excellent agreement with the experimental data for all configurations. 
From the fits, we observed an increase in capacitance and carrier-induced current with larger micro-cell sizes and coated devices. The validated model provides a reliable basis for predicting the IV behaviour of the sensors under irradiation, removing the need for repeated measurements in future studies.

\begin{figure}[t!]
    \centering
    \includegraphics[height=4.8cm]{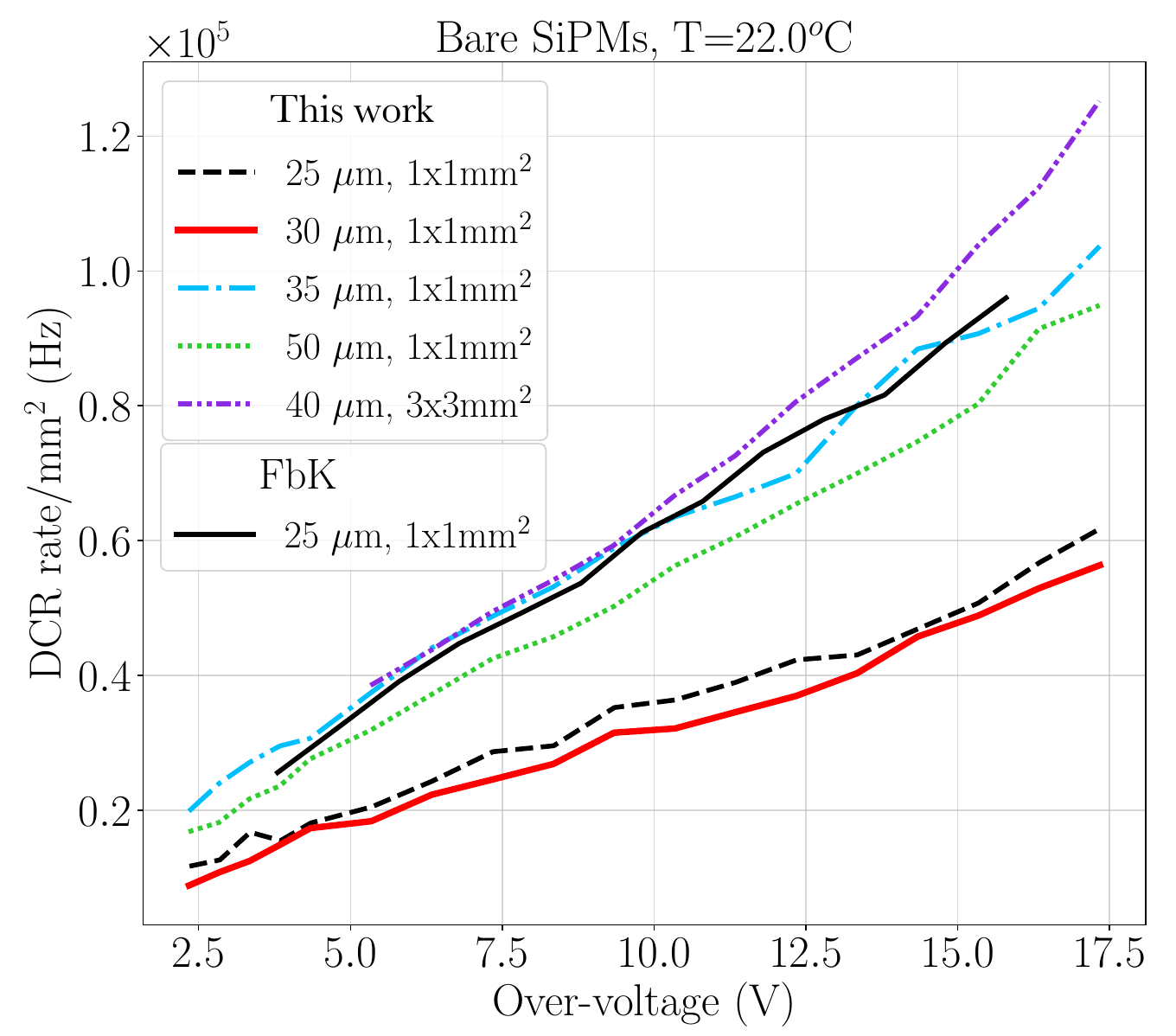}
    \includegraphics[height=4.8cm]{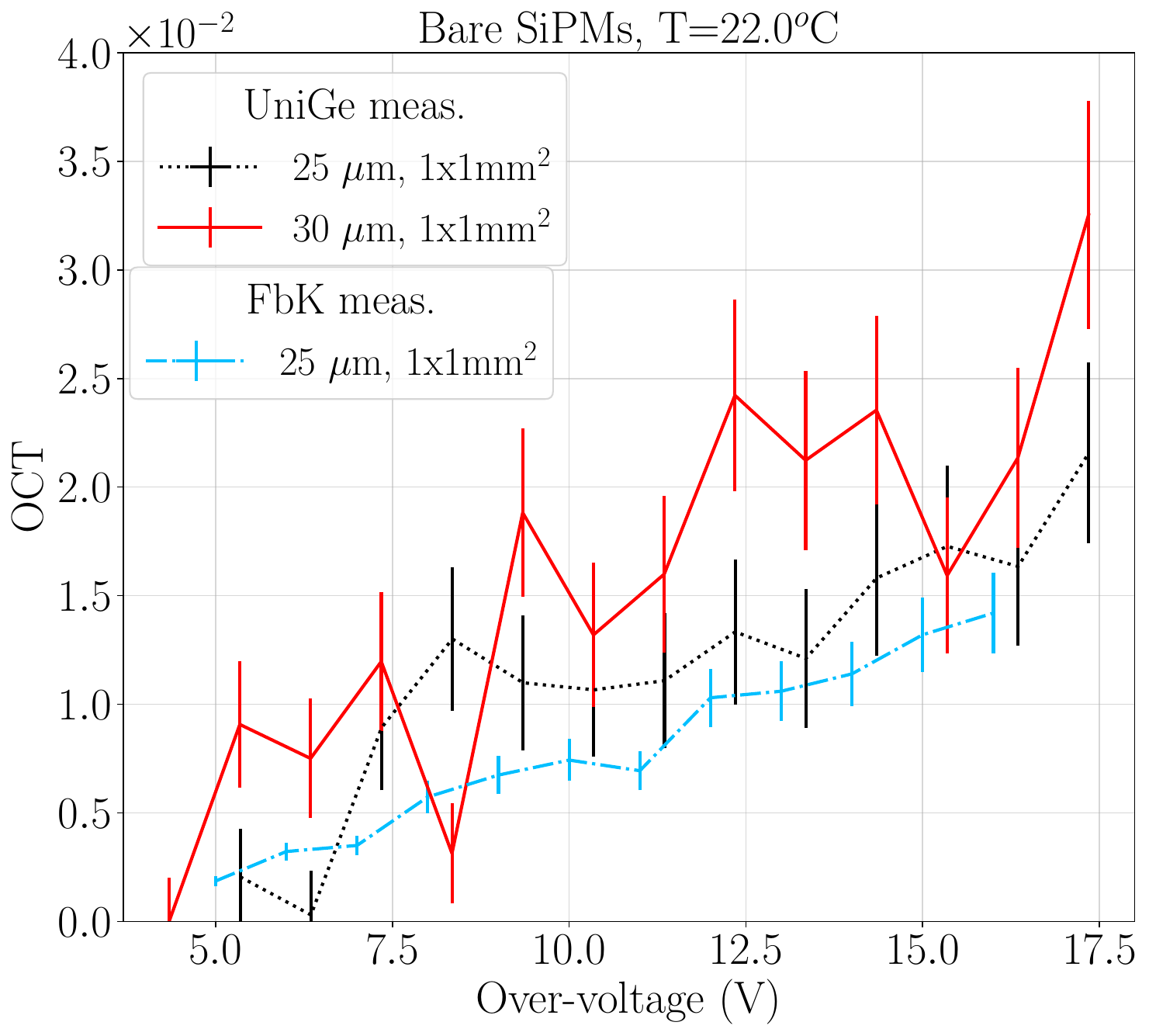}
    \caption{Left panel: Characterisation measurements of DCR per mm$^2$ as a function of the overvoltage for bare sensors with different micro-cell sizes compared to FBK's measurements. Right panel: Our and FBK measurements of OCT as a function of overvoltage for sensors of 25 and 30~$\mu$m micro-cell sizes without coating.}
    \label{fig:dcr_oct}
\end{figure}

%\subsection*{Dynamic Response and Noise Analysis} % Dynamic characterization
\textbf{Dynamic characterization:} We performed the dynamic characterisation of the SiPMs by measuring primary uncorrelated noise (i.e. DCR), and secondary correlated noise components such as optical crosstalk (OCT). Using a custom-built amplifier and waveform acquisition system, we conducted an overvoltage scan from 1 to 18~V and collected 10,000 waveforms per setting. 
%The DCR was modeled as a thermally driven process influenced by electric field strength, while OCT and AP were attributed to secondary avalanches from hot carrier luminescence and carrier trapping, respectively. 
A typical single photoelectron signal measured with a custom pre-amplifying stage (different from the final readout ASIC) is shown in Fig.~\ref{fig:IV_postbreakdown_fit}. The DCR and OCT results are presented in Fig.~\ref{fig:dcr_oct}, where we observe that at 10V overvoltage, the DCR remains below 60~kHz/mm$^2$ and the OCT below 2\% for micro-cell sizes of 25 and 30~$\mu$m. These results are consistent with FBK reference data despite our setup’s sensitivity to residual environmental noise, which we mitigated using a Faraday cage.

\begin{figure}[t!]
    \centering
    \includegraphics[height=4.5cm]{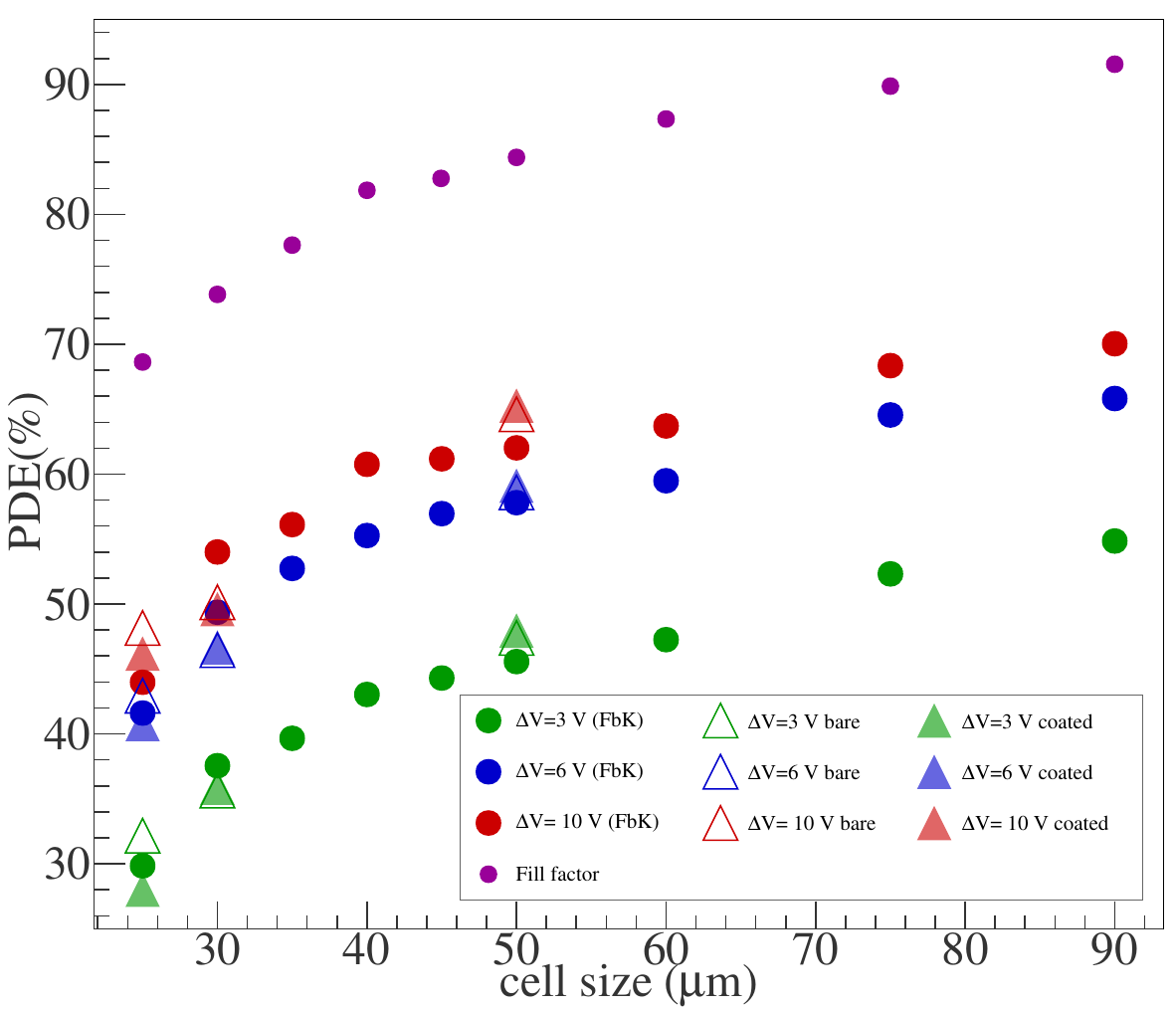}
    \includegraphics[height=5cm]{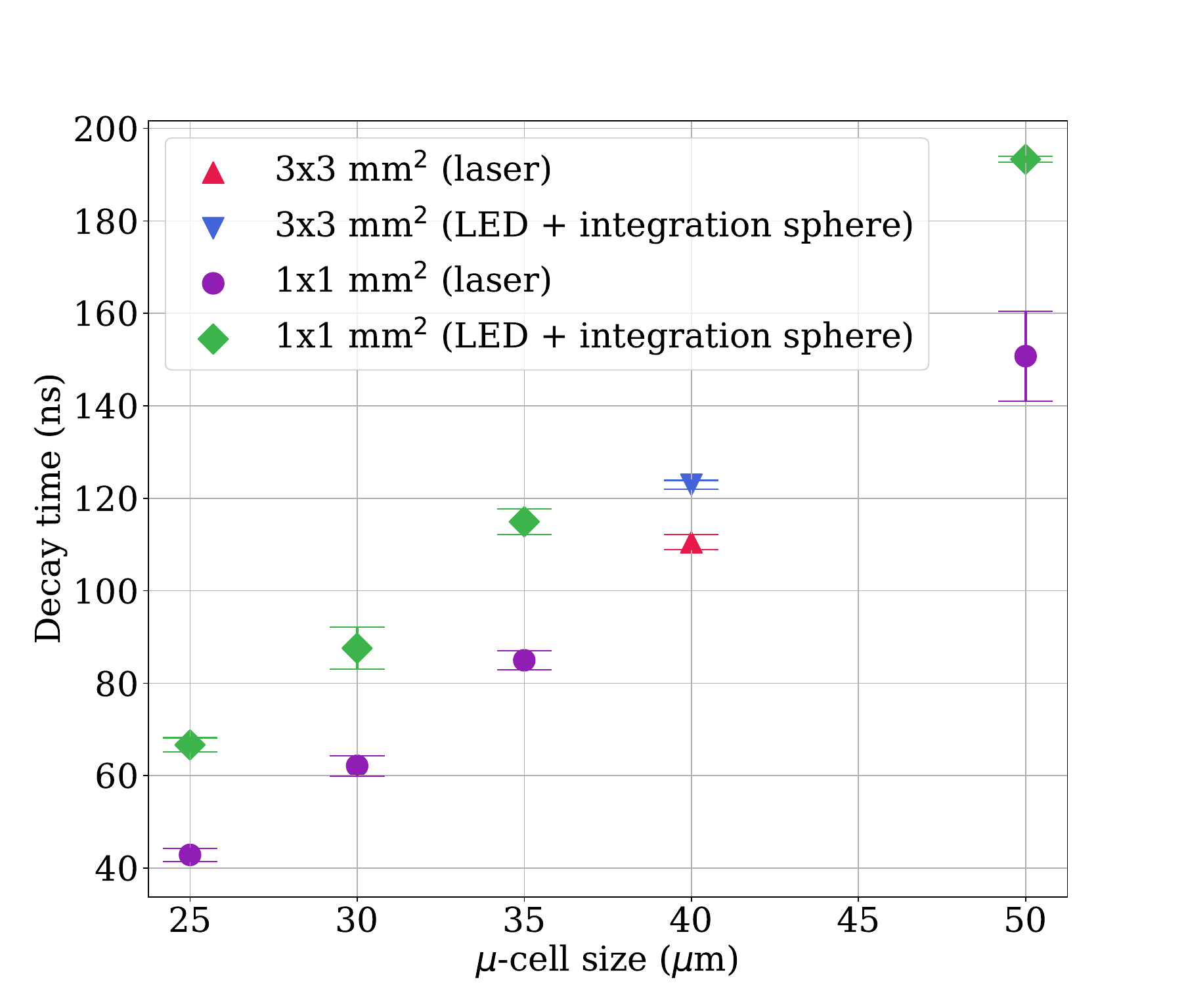}
    \caption{Left panel: PDE vs cell size for different overvoltages and for bare and coated sensors. In addition to the measurements made in Geneva at 450~nm, the FBK measurements are shown for 420~nm (full circles). Right panel: SiPM signal decay time as a function of the $\mu$-cell size measured with a LED (450~nm) and a short pulsed laser (370~nm), respectively with and without an integration sphere.}
    \label{fig:diff_cell}
\end{figure}

%\subsection*{Spectral and Optical Response Evaluation} % Optical characterization
\textbf{Optical characterization:} We characterised the PDE of SiPMs as a function of wavelength and overvoltage. The PDE is defined by the product of quantum efficiency, fill factor, and Geiger probability, all of which depend on micro-cell geometry and operating conditions. Using a calibrated photodiode and a pulsed LED source at 450~nm, we performed absolute PDE measurements, accounting for uncorrelated noise. Results in Fig.~\ref{fig:diff_cell} show that the 30~$\mu$m bare sensors provide the best trade-off between high PDE and low DCR, achieving a PDE > 50\% at 10~V overvoltage. This configuration also minimizes OCT, confirming its suitability for low-background, space-based detection.

To further evaluate the optimal configuration, we examined the signal decay time, defined as the time required for the signal to drop from 95\% to 5\% of its peak value, using both a pulsed LED and a picosecond laser at 370~nm, with and without an integrating sphere, and without an amplifier in the readout. As shown in Fig.~\ref{fig:diff_cell}, the decay time increases with larger micro-cell sizes due to increased capacitance. The shortest signals were obtained using the laser without the integrating sphere, confirming minimal temporal spread. 
%As expected, no significant variation in decay time was observed with changing bias voltage. 

Based on the combined analysis of DCR, PDE, OCT, and signal duration, we selected the 30~$\mu$m bare SiPMs for the Terzina focal plane, balancing detection efficiency, timing performance, and power consumption constraints critical for space missions.

\section{Radiation Response Studies of SiPMs}
\label{sec:rad_study}
\begin{figure}[t!]
    \centering
    \includegraphics[width=0.48\textwidth]{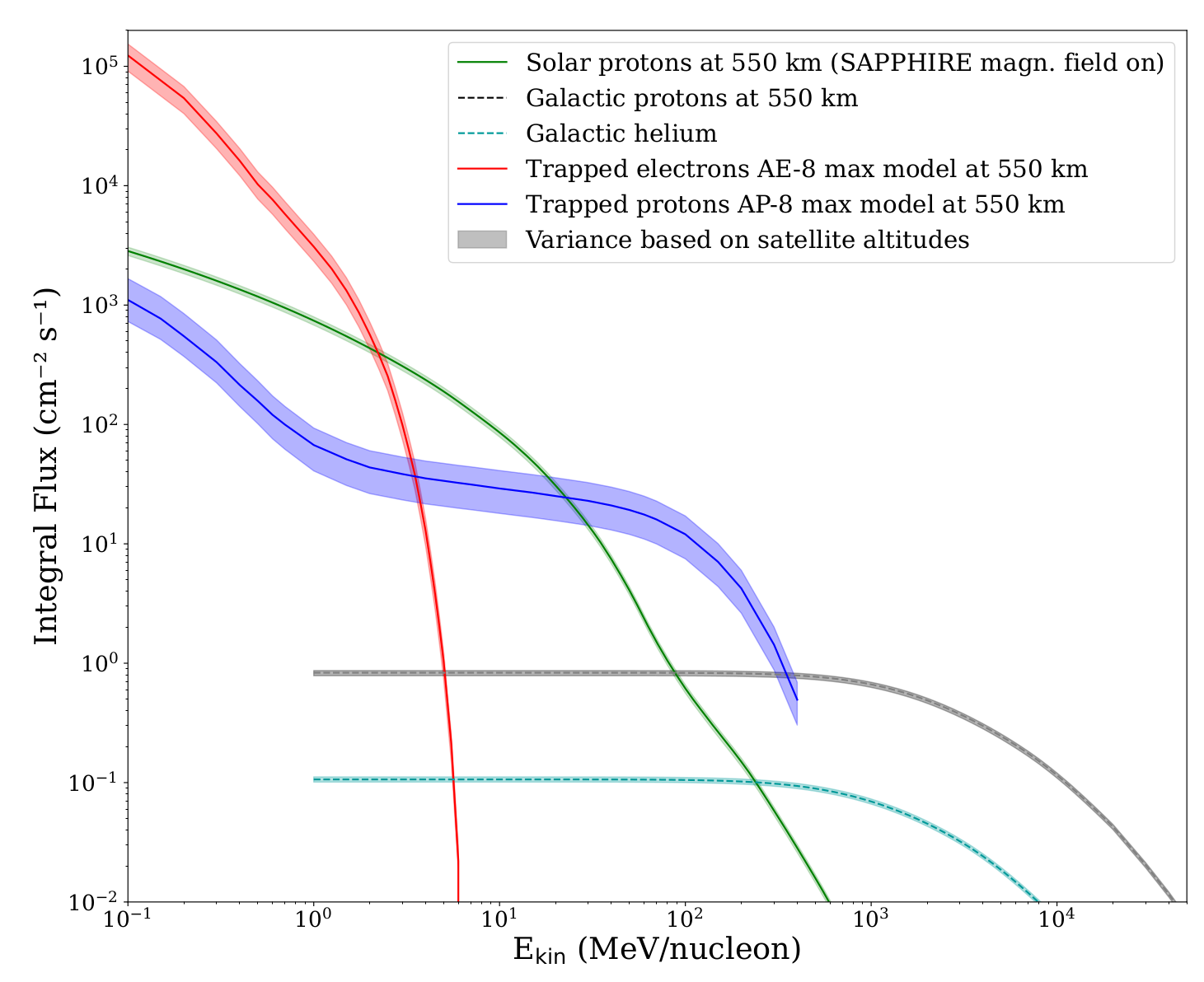}
    \caption{Integral fluxes calculated with SPENVIS for the NUSES mission, showing the expected fluxes of trapped electrons and protons in the Van Allen belts, as well as solar protons, galactic protons, and helium nuclei, at an altitude of 550 km.}
    \label{fig:all_flux_spenvis}
\end{figure}

The Terzina telescope is directly exposed to space radiation due to the absence of atmospheric shielding. Using SPENVIS~\cite{SPENVIS}, we evaluated the radiation environment for the three-year mission duration, 
%(mid-2026 to mid-2029),
identifying trapped protons and electrons in the Van Allen belts, solar protons, and galactic cosmic rays (GCRs) as the primary contributors. As shown in Fig.~\ref{fig:all_flux_spenvis}, the most significant fluxes arise from trapped and solar particles, which can lead to radiation damage. This damage increases the DCR, reducing image quality and effectively raising the trigger threshold.
To mitigate this, Terzina will periodically perform annealing with heating resistors located on the back of SiPM tiles, a method that partially restores the lattice structure and limits DCR increase over time.

% \subsection*{Proton and Electron Irradiation Campaign}
\textbf{Proton and Electron Irradiation Campaign:} We measured the SiPM currents as a function of bias voltage in the post-breakdown Geiger regime under two irradiation conditions. For the \textit{proton irradiation study}, we conducted the test at IFJ PAN in Krakow~\cite{SWAKON20101469} using a 50MeV proton beam. 
% The proton beam spot had a circular shape with a 35~mm diameter and homogeneity better than 5\% with respect to the mean fluence. 
Measurements were taken for $1\times1$ mm$^2$, 30$\mu\text{m}$ cell size SiPMs with resin coating. For the \textit{electron irradiation study}, we used the beta-emitting isotope strontium-90 and measured $3\times3$ mm$^2$, 40~$\mu\text{m}$ cell size SiPMs with resin coating.
%The isotope Strontium~90 with an atomic number of 38 and a half-life of about 28.8 years exclusively undergoes beta decay, with a decay energy of 0.546 MeV, into yttrium-90 and produces an electron and a neutrino. Yttrium-90 (Y90) with a half-life of 64.2~hours also beta decay to stable zirconium (Zr). 
This source enables the study of SiPM radiation damage from electrons with energies up to 2.2~MeV. Considering the fluxes shown in Fig.\ref{fig:all_flux_spenvis}, we use Geant4~\cite{ALLISON2016186} to estimate the expected non-ionizing dose on the Terzina camera, as well as the non-ionizing dose rate for a single SiPM pixel, based on the exact telescope structure. Overall, We obtained 10.1~Gy\footnote{1~rad = 0.01~Gy = 0.01~J/kg}/year for electrons and 2.7~Gy/year for protons.

% \subsection*{Thermal Dependence of SiPM Parameters} % Thermal effect
\textbf{Thermal Dependence of SiPM Parameters:} The optimal operating temperature for Terzina is set at $-20^\circ$C, with thermal control preventing it from exceeding $5^\circ$C. While sensor characterisation was conducted at room temperature, additional IV measurements in a climate chamber were used to study sensor behaviour across a wider range of temperatures. Since DCR is proportional to the post-breakdown current, these temperature-dependent current measurements allowed extrapolation of the DCR under different thermal conditions. The current follows an exponential dependence on temperature, characterised by fit parameters that enable correction to the target operational temperature. Results show that current doubles approximately every $10^\circ$C before irradiation and every $15^\circ$C after. Additionally, the breakdown voltage increases linearly with temperature (by about 30~mV/$^\circ$C on average) while the quenching resistance remains unaffected. These findings support optimal sensor selection and highlight the importance of maintaining stable operation in orbit, which will be achieved by implementing a feedback loop to compensate for temperature-induced shifts in the SiPM working point.

\begin{figure}[t!]
    \centering
    \includegraphics[width=0.8\textwidth]{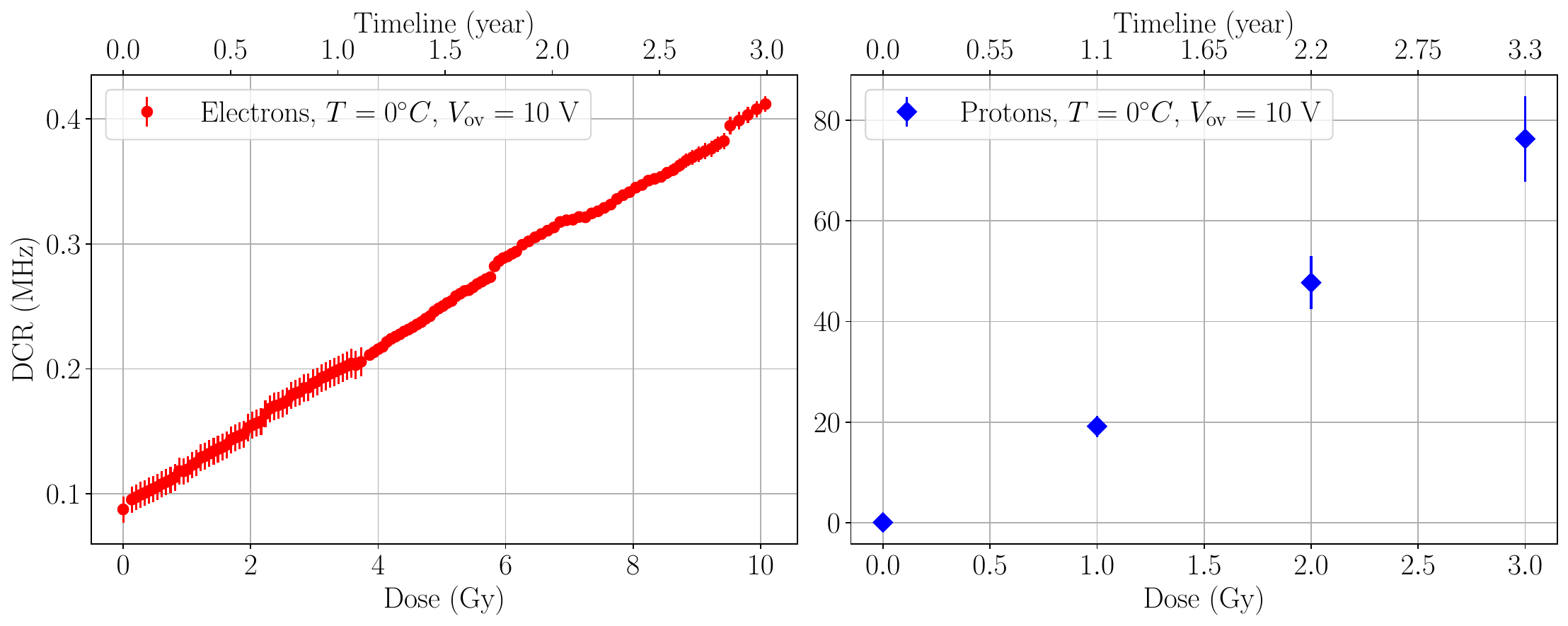}
    \caption{Inferred DCR per pixel of Terzina SiPMs with sensitive area of $\sim 6.58~\text{mm}^2$ from IV measurements during the irradiation as a function of dose and mission timeline. The errors are due to the instrument scale precision.}
    \label{fig5:DCR_proton_electron}
\end{figure}

%\subsection*{Dark Count Rate Forecast for the Mission} % DCR prediction during the mission
\textbf{Dark Count Rate Forecast for the Mission:} The DCR represents the baseline signal rate in the absence of light. To prevent the camera from constantly triggering on noise, the trigger threshold must increase with DCR, which in turn raises the minimum detectable energy of primary cosmic rays. Based on our 
%thermal, electron, and proton irradiation 
measurements, we estimated the DCR accounting for the measured current, correcting for SiPM size and scaling to the operational temperature of $0^\circ$C, as well as the SiPM gain and sensitive area. By combining these results with simulation-based dose rates, we obtained the DCR trends shown in Fig.~\ref{fig5:DCR_proton_electron}. As expected, electron-induced damage is minimal compared to protons: a 2~Gy electron-induced dose increases the DCR to roughly 0.15~MHz, while the same dose from protons leads to a much higher DCR of approximately 50~MHz, assuming identical conditions.

\begin{figure}[t!]
    \centering
    \includegraphics[width=0.55\textwidth]{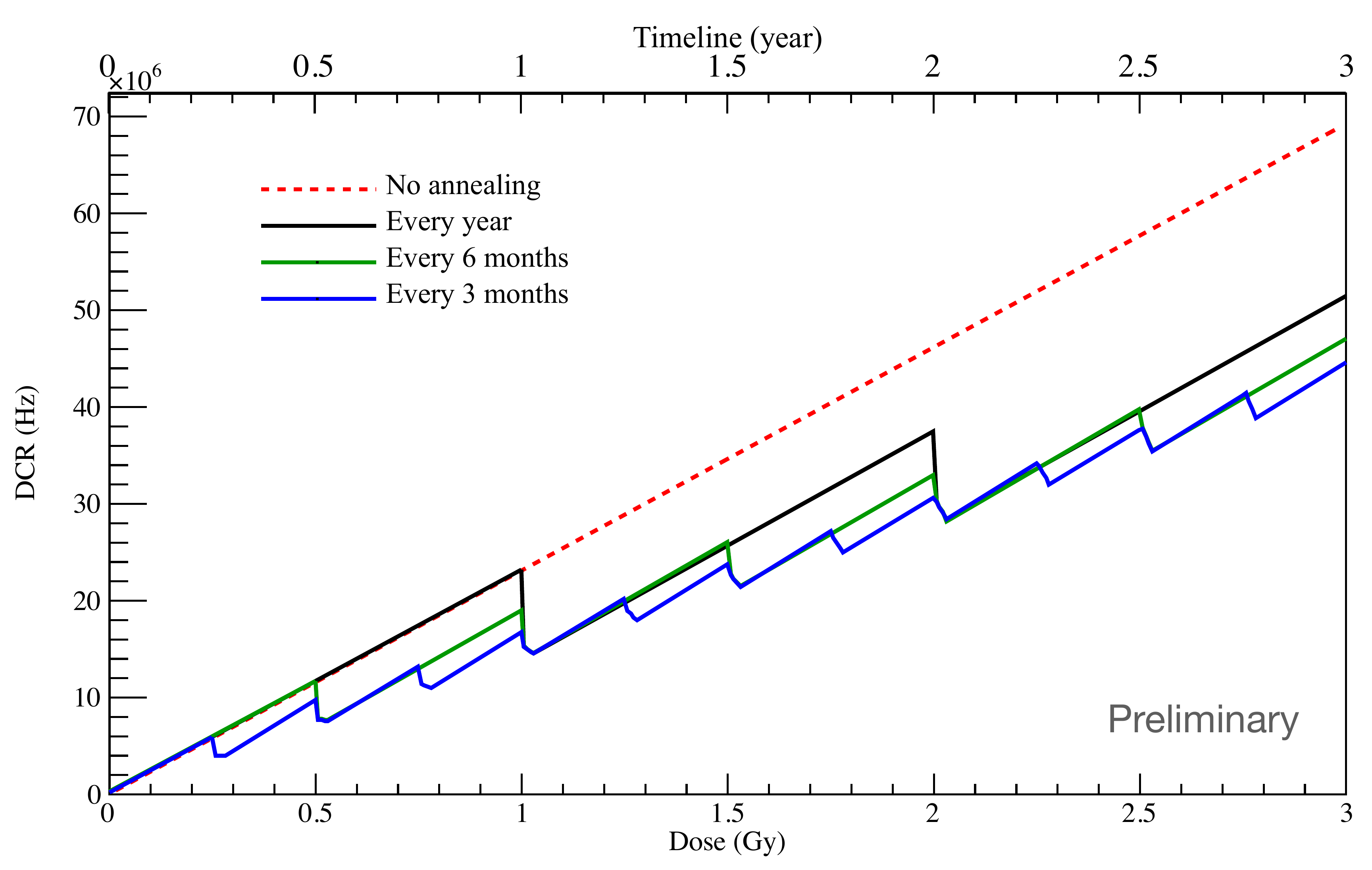}
    \caption{The DCR change from the proton irradiation and annealing cycles during the satellite lifetime.}
    \label{fig6:annealing}
\end{figure}

% \subsection*{Radiation Damage Mitigation via Annealing} % Annealing
\textbf{Radiation Damage Mitigation via Annealing:} To mitigate radiation-induced damage, a thermal annealing strategy was developed and tested. 
%This controlled heating process partially restores SiPM performance by reducing defects and internal stress caused by irradiation. 
Laboratory measurements involved monitoring irradiated sensors over time at fixed annealing temperatures and quantifying recovery through IV curve comparisons. The current, and hence DCR, decreases exponentially with time, with the recovery efficiency depending on the annealing temperature. At temperatures above $50^\circ$C, up to 40\% of the radiation-induced current increase can be recovered over an 84-hour cycle. Based on these results, a predictive model was established to simulate the combined effects of ongoing irradiation and periodic annealing during the mission. As shown in Fig.~\ref{fig6:annealing}, different annealing strategies influence the long-term evolution of the DCR, enabling more stable performance of the SiPMs throughout the satellite’s lifetime.

% \section{Focal Plane Assembly and Tile-Level Performance}
% \label{sec:fpa_perf}

\section{Conclusion}
\label{sec:conclusion}
This work presents the characterisation of FBK SiPMs, which allowed to select the optimal size of the $\mu$-cells of SiPMs to adopt in the Terzina mission, and the study that determines the power consumption increase with radiation damage. It also lead to the definition of a mitigation strategy based on periodic annealing during the mission.
A model has been developed from the measurements which allows other users to determine similar mitigation strategies to this relevant damaging effect of SiPMs.

\section*{Acknowledgments}
NUSES is funded by the Italian Government (CIPE n. 20/2019), by the Italian Ministry of Economic Development (MISE reg. CC n. 769/2020), by the Italian Space Agency (CDA ASI n. 15/2022), 
by the European Union NextGenerationEU under the MUR National Innovation Ecosystem grant ECS00000041 - VITALITY - CUP D13C21000430001 and by the Swiss National Foundation (SNF grant n. 178918). 
This study was carried out also in collaboration with the Ministry of University and Research under contract n. 2024-5-E.0 - CUP n. I53D24000060005. 
This research, leading to the beam test results, also received partial funding from the European Union’s Horizon Europe research and innovation programme under grant agreement No. 101057511.

\begingroup
\renewcommand*{\bibfont}{\footnotesize}

\endgroup

\begin{center}
{
\large 
\bf
The NUSES Collaboration 
}
\vspace{5ex}

M.~Abdullahi$^{a,b}$, R.~Aloisio$^{a,b}$, F.~Arneodo$^{c,d}$, S.~Ashurov$^{a,b}$, U.~Atalay$^{a,b}$, F.~C.~T.~Barbato$^{a,b}$, R.~Battiston$^{e,f}$, M.~Bertaina$^{g,h}$, E.~Bissaldi$^{i,j}$, D.~Boncioli$^{k,b}$, L.~Burmistrov$^{l}$, F.~Cadoux$^{l}$, I.~Cagnoli$^{a,b}$, E.~Casilli$^{a,b}$, D.~Cortis$^{b}$, A.~Cummings$^{m}$, M.~D'Arco$^{l}$, S.~Davarpanah$^{l}$, I.~De~Mitri$^{a,b}$, G.~De~Robertis$^{i}$, A.~Di~Giovanni$^{a,b}$, A.~Di~Salvo$^{h}$, L.~Di~Venere$^{i}$, J.~Eser$^{n}$, Y.~Favre$^{l}$, S.~Fogliacco$^{a,b}$, G.~Fontanella$^{a,b}$, P.~Fusco$^{i,j}$, S.~Garbolino$^{h}$, F.~Gargano$^{i}$, M.~Giliberti$^{i,j}$, F.~Guarino$^{o,p}$, M.~Heller$^{l}$, T.~Ibrayev$^{c,d,q}$, R.~Iuppa$^{e,f}$, A.~Knyazev$^{c,d}$, J.~F.~Krizmanic$^{r}$, D.~Kyratzis$^{a,b}$, F.~Licciulli$^{i}$, A.~Liguori$^{i,j}$, F.~Loparco$^{i,j}$, L.~Lorusso$^{i,j}$, M.~Mariotti$^{s,t}$, M.~N.~Mazziotta$^{i}$, M.~Mese$^{o,p}$, M.~Mignone$^{g,h}$, T.~Montaruli$^{l}$, R.~Nicolaidis$^{e,f}$, F.~Nozzoli$^{e,f}$, A.~Olinto$^{u}$, D.~Orlandi$^{b}$, G.~Osteria$^{o}$, P.~A.~Palmieri$^{g,h}$, B.~Panico$^{o,p}$, G.~Panzarini$^{i,j}$, D.~Pattanaik$^{a,b}$, L.~Perrone$^{v,w}$, H.~Pessoa~Lima$^{a,b}$, R.~Pillera$^{i,j}$, R.~Rando$^{s,t}$, A.~Rivetti$^{h}$, V.~Rizi$^{k,b}$, A.~Roy$^{a,b}$, F.~Salamida$^{k,b}$, R.~Sarkar$^{a,b}$, P.~Savina$^{a,b}$, V.~Scherini$^{v,w}$, V.~Scotti$^{o,p}$, D.~Serini$^{i}$, D.~Shledewitz$^{e,f}$, I.~Siddique$^{a,b}$, L.~Silveri$^{c,d}$, A.~Smirnov$^{a,b}$, R.~A.~Torres~Saavedra$^{a,b}$, C.~Trimarelli$^{a,b}$, P.~Zuccon$^{e,f}$, S.~C.~Zugravel$^{h}$.

 \vspace{5ex}

\begin{tabular}{c}
$^{a}$ Gran Sasso Science Institute (GSSI);\\ 
$^{b}$ Istituto Nazionale di Fisica Nucleare (INFN) - Laboratori Nazionali del Gran Sasso;\\ 
$^{c}$ Center for Astrophysics and Space Science (CASS);\\ 
$^{d}$ New York University Abu Dhabi, UAE;\\ 
$^{e}$ Dipartimento di Fisica - Università di Trento;\\ 
$^{f}$ Istituto Nazionale di Fisica Nucleare (INFN) - Sezione di Trento;\\ 
$^{g}$ Dipartimento di Fisica - Università di Torino;\\ 
$^{h}$ Istituto Nazionale di Fisica Nucleare (INFN) - Sezione di Torino;\\ 
$^{i}$ Istituto Nazionale di Fisica Nucleare (INFN) - Sezione di Bari;\\ 
$^{j}$ Dipartimento di Fisica M. Merlin dell’ Università e del Politecnico di Bari;\\ 
$^{k}$ Dipartimento di Scienze Fisiche e Chimiche -Università degli Studi di L’Aquila;\\ 
$^{l}$ Départment de Physique Nuclèaire et Corpuscolaire - Université de Genève, Faculté de Science;\\ 
$^{m}$ Department of Physics and Astronomy and Astrophysics, Institute for Gravitation and the Cosmos;\\ 
$^{n}$ Department of Astronomy and Astrophysics, University of Columbia;\\ 
$^{o}$ Istituto Nazionale di Fisica Nucleare (INFN) - Sezione di Napoli;\\ 
$^{p}$ Dipartimento di Fisica E. Pancini - Università di Napoli Federico II;\\ 
$^{q}$ now at The School of Physics, The University of Sydney;\\ 
$^{r}$ CRESST/NASA Goddard Space Flight Center;\\ 
$^{s}$ Dipartimento di Fisica e Astronomia - Università di Padova;\\ 
$^{t}$ Istituto Nazionale di Fisica Nucleare (INFN) - Sezione di Padova;\\ 
$^{u}$ Columbia University, Columbia Astrophysics Laboratory;\\ 
$^{v}$ Dipartimento di Matematica e Fisica “E. De Giorgi” - Università del Salento;\\ 
$^{w}$ Istituto Nazionale di Fisica Nucleare (INFN) - Sezione di Lecce. \\
\end{tabular}

\end{center}

\end{document}